\documentclass[aps,twocolumn,showpacs]{revtex4}
\usepackage{amsmath}
\usepackage{bm}
\usepackage{mathrsfs}
\usepackage{mathptmx}

\begin{document}

\title{Spin Solitons in Magnetized Pair Plasmas}
\author{G. Brodin and M. Marklund}
\affiliation{Department of Physics, Ume{\aa} University, SE--901 87 Ume{\aa},
Sweden}

\date{{\normalsize Submitted to Physics of Plasmas, July 31, 2007;
  resubmitted September 4, 2007; accepted September 12, 2007}}

\begin{abstract}
A set of fluid equations, taking into account the spin properties of the
electrons and positrons in a magnetoplasma, are derived. The
magnetohydrodynamic limit of the pair plasma is investigated. It is shown
that the microscopic spin properties of the electrons and positrons can lead
to interesting macroscopic and collective effects in strongly magnetized
plasmas. In particular, it is found that new Alfv\'{e}nic solitary
structures, governed by a modified Korteweg--de Vries equation, are allowed
in such plasmas. These solitary structures vanish if the quantum spin
effects are neglected. Our results should be of relevance for astrophysical
plasmas, e.g. in pulsar magnetospheres, as well as low-temperature laboratory plasmas. 
\end{abstract}

\maketitle



\section{Introduction}


The magnetoplasma, first discussed by Alfv\'en \cite{alfven}, is today a
mature research topic, with a wide range of applications. Recently, so
called quantum plasmas in which the quantum properties of the plasma
particles are taken into account, have received attention (see e.g. \cite
{manfredi,marklund-brodin,brodin-marklund} for an up-to-date set of
references). The collective motion of quantum particles in magnetic fields
thus gives a natural extension to the classical theory of
magnetohydrodynamics (MHD) in terms of so called quantum magnetoplasmas. The
discussion of quantum plasmas has shown that quantum collective effects can
have interesting consequences both in laboratory and astrophysical
environments \cite{haas-etal1,anderson-etal,haas-etal2,haas,garcia-etal,marklund,shukla-stenflo,shukla,Shukla-Eliasson,shukla-etal,melrose,melrose-weise,baring-etal,harding-lai}, and part of the
literature has indeed been influenced by experimental progress and
techniques (see e.g. \cite{marklund-brodin-stenflo,marklund-shukla,exp1,exp2,lundstrom-etal,lundin-etal,dipiazza-etal}, where also
strong field effects are discussed). From the laboratory perspective, the
motion of particles with quantum spin in strong fields, using e.g. intense
lasers, has attracted interest as a probe of quantum physical phenomena \cite
{rathe-etal,hu-keitel,arvieu-etal,aldana-roso,walser-keitel,walser-etal}.
These studies are however mainly focused on single particle properties. In
Refs.\ \cite{cowley-etal,kulsrud-etal} however, the kinetic properties of
spin plasmas was investigated. Strong fields appear in pulsar and magnetar
environments \cite{Beskin-book,asseo,magnetar}. Discussions of quantum
plasmas in such environments can be found in Refs.\ \cite
{melrose,melrose-weise,baring-etal,harding-lai}. Moreover, studies taking
both certain quantum electrodynamical effects, such as photon splitting, as
well as collective particle effects have been made \cite
{Brodin-etal-2006,Brodin-etal-2007}.

In the present paper, starting from the basic set of equations presented in
Refs.\ \cite{marklund-brodin} and \cite{brodin-marklund}, we derive the
governing MHD equations for a pair plasma. These MHD equations takes into
account quantum properties, such as spin, of the electrons and positrons. It
is found that the nonlinear propagation of Alfv\'en waves in such quantum
MHD (QMHD) pair plasmas is governed by a modified Korteweg--de Vries
equation. This equation is known to support solitary structures. Moreover,
if the quantum spin effects are neglected, the solitary structures vanish,
making them true quantum solitons. The results should be of interest for
both laboratory and astrophysical plasmas.


\section{Governing equations}


In Refs.\ \cite{marklund-brodin} and \cite{brodin-marklund}, it was shown
that the evolution of spin plasma particles in a strongly magnetized
environment is governed by the equations for the particle densities $n_{j}$
and velocities $\mathbf{V}_{j}$ 
\begin{equation}
\frac{\partial n_{j}}{\partial t}+\mathbf{\nabla}\cdot (n_{j}\mathbf{V}_{j})=0,
\label{eq:density}
\end{equation}
and 
\begin{equation}
mn_{j}\left( \frac{\partial }{\partial t}+\mathbf{V}_{j}\cdot \mathbf{\nabla}\right) %
\mathbf{V}_{j}=q_{j}n_{j}\left( \mathbf{E}+\mathbf{V}_{j}\times \mathbf{B}\right) -%
\mathbf{\nabla}P_{j}+\mathbf{\mathcal{C}}_{j}+\mathbf {F}_{Q,j},  \label{eq:mom-q}
\end{equation}
where $q_{j}$ stands for the charge of the electrons ($q_{e}=-e$) or
positrons ($q_{p}=e$), $m=m_{e}=m_{p}$ is the electron (positron) mass, $%
\mathbf
{\mathcal{C}}_{j}$ is the collisional contribution between species $j$ and
the second species, and $\mathbf{F}_{Q,j}$ is the total quantum force density
(for a complete expression of this force density, see Refs.\ \cite
{marklund-brodin} and \cite{brodin-marklund}). Here we will make use of the
following expression for the quantum force on the electrons/positrons 
\begin{equation}
\mathbf{F}_{Q,j}=n_{j}\left[ \mathbf{\nabla}\left( \frac{\hbar ^{2}}{2mn_{j}^{1/2}}%
\nabla ^{2}n_{j}^{1/2}\right) + \tanh \left( \frac{\mu _{B}B}{T_{j}}\right)
\mu _{B}\mathbf{\nabla}B\right] ,  \label{eq:f-closed}
\end{equation}
where the first term is the gradient of the so called Bohm potential, the
second term comes from the spin and $B=\left| \mathbf{B}\right| $. We note that $%
\tanh (x)=B_{1/2}(x)$, where $B_{1/2}$ is the Brillouin function with
argument $1/2$ describing particles of spin 1/2. The temperature $T_{j}$ is
measured in units of energy. Furthermore, we have introduced the Bohr
magneton $\mu _{B}=e\hbar /2m$, where $\hbar $ is Planck's constant divided
by $2\pi $.

The coupling between the quantum plasma species is mediated by the
electromagnetic field. The total magnetic field include both the classical
contribution (from currents $\mathbf{j}=\sum q_{j}n_{j}\mathbf{V}_{j}$) and the spin
sources, such that Amp\`{e}re's law reads 
\begin{equation}
\mathbf{\nabla}\times \mathbf{B}=\mu _{0}(\mathbf{j}+\mathbf{j}_M)+\frac{1}{c^{2}}\frac{%
\partial \mathbf{E}}{\partial t},  \label{Eq-ampere}
\end{equation}
including the magnetization spin current $\mathbf{j}_{M,j}=\mathbf{\nabla}\times
(2q_{j}n_{j}\mu _{B}\mathbf{S}/\hbar \left| q_{j}\right| ) $ for each species,
where $\mathbf{S}$ is the spin vector. Furthermore, we need Faraday's law 
\begin{equation}
\mathbf{\nabla}\times \mathbf{E}=-\frac{\partial \mathbf{B}}{\partial t}.
\label{Eq-Faraday}
\end{equation}
We note that the spin current is determined by the spin vector $\mathbf{S}$. In
general the spin vector follows a separate evolution equation, see Ref. \cite
{marklund-brodin}, but in the limit where the time-scales are much longer
than the Larmor period, the spin vector can be approximated by 
\begin{equation*}
\mathbf{S}=\frac{\hbar }{2}\frac{q_{j}}{\left| q_{j}\right| }\tanh \left( \frac{%
\mu _{B}B}{T_{j}}\right) \mathbf{\hat{B}}
\end{equation*}
where $\mathbf{\hat{B}}$ is a unit vector in the direction of the magnetic
field. The magnetization spin current is thus given by 
\begin{equation}
\mathbf{j}_M=\sum_{j}\mathbf{\nabla}\times \left[ n_{j}\mu _{B}\tanh \left( \frac{%
\mu _{B}B}{T_{j}}\right) \mathbf{\hat{B}}\right]  \label{Eq-spin-current-1}
\end{equation}

\section{Electron--positron plasma and the MHD limit}


We now introduce the total mass density $\rho \equiv m(n_{e}+n_{p})$, the
centre-of-mass fluid flow velocity $\mathbf{V}\equiv m(n_{e}\mathbf{V}_{e}+n_{p}%
\mathbf{V}_{p})/\rho $, and the current density $\mathbf{j}=-en_{e}\mathbf{V}_{e}+en_{p}%
\mathbf{V}_{p}$, and assume that $P_{e,p}=T_{e,p}n_{e,p}$. Using these
definitions, we immediately obtain 
\begin{equation}
\frac{\partial \rho }{\partial t}+\mathbf{\nabla}\cdot (\rho \mathbf{V})=0.
\label{eq:mhd-cont}
\end{equation}
Assuming quasi-neutrality, i.e.\ $n_{e}\approx n_{p}$, we next add the
momentum conservation equations for the electrons and positrons to obtain 
\begin{equation}
\rho \left( \frac{\partial }{\partial t}+\mathbf{V}\cdot \mathbf{\nabla}\right) %
\mathbf{V}=\mathbf{j}\times \mathbf{B}-\mathbf{\nabla}\cdot \mathbf{\mathsf{\Pi}}+\mathbf{F}_{Q},
\label{eq:mhd-mom}
\end{equation}
where $\mathbf{\mathsf{\Pi}}=[(T_{e}+T_{p})/2m_{e}]\mathbf{\mathsf{I}}+(m^{2}/\rho )%
\mathbf{j}\otimes \mathbf{j}$ is the total pressure tensor in the centre-of-mass
frame and 
\begin{eqnarray}
  \mathbf{F}_{Q}=\mathbf{F}_{Q,e}+\mathbf{F}_{Q,p}=&&\!\!\!\!\!\!\!\! \rho \Bigg[ \mathbf{\nabla}\left( \frac{%
  \hbar ^{2}}{2m^{2}\rho ^{1/2}}\nabla ^{2}\rho ^{1/2}\right) 
\nonumber \\ && \quad
  + \sum_{j}\tanh
  \left( \frac{\mu _{B}B}{T_{j}}\right) \frac{\mu _{B}}{m}\mathbf{\nabla}B\Bigg] .
\label{eq:quantum2}
\end{eqnarray}
The collisional contributions have here cancelled due to momentum
conservation. Subtracting the momentum equations for the electrons and
positrons, assuming $\mathbf{\mathcal{C}}_{e}=\eta en_{e}\mathbf{j}$, where $\eta $
is the resistivity, we also obtain the generalized Ohm's law 
\begin{eqnarray}
  \frac{\partial \mathbf{j}}{\partial t}+\mathbf{\nabla}\cdot \left( \mathbf{j}\otimes %
  \mathbf{V}+\mathbf{V}\otimes \mathbf{j}\right) =&&\!\!\!\!\!\!\!\!  \frac{e^{2}\rho }{m}\left( \mathbf{E}+\mathbf{V}%
  \times \mathbf{B}-\eta \mathbf{j}\right) 
\nonumber \\ &&\!\!\!\!\!\!\!\!
  -\frac{e}{2m^{2}}(\Delta T)\mathbf{\nabla}\rho -%
  \frac{e}{m}\Delta \mathbf{F}_{Q},  \label{eq:general-dynamo}
\end{eqnarray}
where $\Delta T=T_{e}-T_{p}$, and $\Delta \mathbf{F}_{Q}=\mathbf{F}_{Q,e}-\mathbf {F}%
_{Q,p}$. We note that in the limit of quasi-neutrality, the non-spin quantum
force contributions to Ohm's law cancel, thus only leaving the spin quantum
force, as long as the species have different temperatures. Eq. (\ref
{eq:general-dynamo}) is valid during rather general conditions, and accounts
for different temperatures of the species, as well as for short scale
lengths of the order of the gyro-radius. From now on we will however limit
ourselves to an equal temperature plasma, with scale lengths longer than the
gyro-radius. Dividing by the density and taking the curl of Eq. (\ref
{eq:general-dynamo}), we obtain 
\begin{equation}
\frac{\partial \mathbf{B}}{\partial t}=\mathbf{\nabla}\times \left( \mathbf{V}\times %
\mathbf{ B}-\eta \mathbf{j}\right)  \label{eq:simple-ohm}
\end{equation}
For the cases where the Alfv\'{e}n speed is much smaller than the speed of
light, the displacement current in (\ref{Eq-ampere}) can be neglected, and
thus the current can be expressed in terms of the magnetic field. In this
case, Eqs.\ (\ref{eq:mhd-cont}), (\ref{eq:mhd-mom}) and (\ref{eq:simple-ohm}%
) constitute a closed system for $\mathbf{B}$, $\mathbf{V}$ and $\rho $. \ But in
plasmas with strong magnetic fields, for example close to pulsars and
magnetars, the Alfv\'{e}n speed approaches $c$, and the displacement current
must be included in Amp\`ere's law. However, we still obtain a closed system
for the same equations (\ref{eq:mhd-cont}), (\ref{eq:mhd-mom}) and (\ref
{eq:simple-ohm}), simply by eliminating the electric field using Ohm's law
and writing the current as 
\begin{equation}
\mathbf{j} = \frac{1}{\mu _{0}}\nabla \times \mathbf{B} - \varepsilon _{0}\frac{%
\partial }{\partial t}\left( \mathbf{V}\times \mathbf{ B}\right) - \mathbf{j}_M
\label{eq:Ampere-Ohm}
\end{equation}
where from (\ref{Eq-spin-current-1}) we write the spin current as 
\begin{equation}
\mathbf{j}_M = \frac{\mu _{B}}{2m}\mathbf{\nabla}\times \left[ \rho \tanh \left( 
\frac{\mu _{B}B}{T}\right) \mathbf{\hat{B}}\right]
\end{equation}
for a quasineutral electron-positron plasma with equal temperature $T$. Note
that in (\ref{eq:Ampere-Ohm}) we have for simplicity omitted dissipative
effects ($\eta \rightarrow 0$).


\section{One-dimensional Alfv\'{e}n waves}


The spin terms have rather different properties than the standard MHD-
terms. As a consequence, the spin force does not necessarily need to be as
large as the ordinary $\mathbf{j}\times \mathbf{B}$-force, in order to significantly
affect the evolution of the system. In order to illustrate this specific
property, we consider the weakly nonlinear evolution of shear Alfv\'{e}n
waves propagating at an angle to the magnetic field. In the absence of spin
effects, the Alfv\'{e}n waves propagates with the Alfv\'{e}n speed $c_{A}=%
\left[ c^{2}B_{0}^{2}/(c^{2}\mu _{0}\rho _{0}+B_{0}^{2})\right] ^{1/2}$,
where $B_{0}$ is the magnitude of the unperturbed magnetic field, and $\rho
_{0}$ is the unperturbed density. We consider the case with spatial
dependence on a single coordinate $\xi =x\sin \theta +z\cos \theta $. As a
prerequisite, we first consider the linearized equations. Letting the
unperturbed magnetic field lie along the $z$-axis, and combining Eqs. (\ref
{eq:mhd-mom}), (\ref{eq:simple-ohm}) and (\ref{eq:Ampere-Ohm}) (with $\eta
\longrightarrow 0$ ) we, as a first approximation, obtain 
\begin{equation}
\left[ \frac{\partial ^{2}}{\partial t^{2}}-c_{A,\mathrm{sp}}^{2}\cos
^{2}\theta \frac{\partial ^{2}}{\partial \xi ^{2}}\right] v_{y}=0
\label{eq:Alf-linear}
\end{equation}
Here $c_{A,\mathrm{sp}}=c_{A}^{2}/(1+\delta _{\rm sp})$ is the spin-modified
Alfv\'{e}n velocity, with the spin modification determined by $\delta _{%
\mathrm{sp}}=\hbar \omega _{p}^{2}\tanh (\mu _{B}B_{0}/T)/2mc^{2}\omega _{c}$, 
where $\omega_p =[(e^2(n_{0p} + n_{0e})/\epsilon_0m]^{1/2}$ is the plasma frequency of the pair plasma, which comes from the part of the spin current proportional to $(\rho_0/m)
\mathbf{\nabla}\times \lbrack \tanh \left( \mu _{B}B_{0}/T\right) \mathbf{\hat{B}}]$%
. Here we have introduced the cyclotron frequency $\omega _{c}=eB_{0}/m$ 
\footnote{In the case when the contribution to external
magnetic field from the spins is large, it is strictly speaking only the
external sources (i.e. not the spin sources) that contribute to the magnetic
field and thereby the cyclotron frequency in this formula. Thus formally we
should replace the full cyclotron frequency $\omega _{c}$ with the cyclotron
frequency due to external sources only, $\omega _{c-\mathrm{ext}}$, where
the two are related by $\omega _{c}=\omega _{c-\mathrm{ext}}+\hbar \omega
_{p}^{2}\tanh (\mu _{B}B_{0}/T)/mc^{2}$. In our case, however, this
difference is negligible.}. For the sake of definiteness we also consider waves
propagating in the positive $\xi $-direction, such that we can use the
expression $\partial /\partial t=c_{A,\mathrm{sp}}\cos \theta \,\partial
/\partial \xi $ in the linear nondispersive approximation. Furthermore, from
now on we assume that the spin effects are small in the sense that $\delta _{%
\mathrm{sp}}\ll 1$, and also that the spins are weakly aligned with the
external magnetic field such that we can approximate $\tanh \left( \mu
_{B}B_{0}/T\right) \approx \mu _{B}B_{0}/T$.

As is wellknown, a weakly nonlinear evolution typically leads to wave
steepening effects, and the necessity to include dispersion on an equal
footing. Following Ref.\ \cite{Brodin-JPP}, but with the inclusion of spin
terms, treating the Hall current, i.e.\ the first term of equation (\ref
{eq:general-dynamo}), as a small correction, weakly dispersive effects are
thus kept. Considering only the positive propagating waves, the linear wave
operator for shear Alfv\'{e}n waves can then be generalized to 
\begin{equation}
\left[ \frac{\partial }{\partial t} + c_{A,\mathrm{sp}}\cos\theta\, \frac{%
\partial }{\partial \xi } + \frac{c_{A}^{3}\cos^3\theta}{2 \omega
_{c}^{2}\sin^2\theta}\frac{\partial ^{3}}{\partial \xi ^{3}}\right] v_{y}=0
\label{eq:Alf-lin-disp}
\end{equation}
where small spin-contribution to the dispersive term in Eq.\ (\ref
{eq:Alf-lin-disp}) has been neglected.

Next, including nonlinear terms, we note that the lowest order terms (i.e.
those proportional to $B_{y}^{2}$ and $v_{y}^{2}$) do not contribute
directly to the nonlinear coupling, and thus we must include nonlinear terms
of higher order in both the amplitude and the $1/\omega _{c}$ expansion. We
thereby obtain 
\begin{eqnarray}
  &&\!\!\!\!\!\!\!\!\!\!\!\!\!\!\!\!\!\!\!\!\!\!\!\!\!\!\!\!\!
  \left[ \frac{\partial }{\partial t}+c_{A,\mathrm{sp}}\cos \theta \,\frac{%
  \partial }{\partial \xi }+\frac{c_{A}^{3}\cos ^{3}\theta }{2\omega
  _{c}^{2}\sin ^{2}\theta }\frac{\partial ^{3}}{\partial \xi ^{3}}\right]
  v_{y}
  \nonumber \\ &&\!\!\!\!\!\!\!\!\!\!\!\!\!\!\!\!\!\!\!\!\!\!\!\!\!
  =\frac{c_{A}\cos \theta }{2\omega _{c}}\frac{\partial }{\partial \xi }%
  \left[ \frac{v_{y}}{\sin \theta }\frac{\partial v_{y}}{\partial \xi }+\omega
  _{c}\frac{v_{y}\rho _{1}}{\rho _{0}}\right] - 
  \frac{(\mu _{B}B_{0})^2}{2mT}\frac{v_{y}^{2}}{c_{A}^{3}}\frac{\partial v_{y}}{%
  \partial \xi }  .
\label{eq:Alf-NL1}
\end{eqnarray}
The linear relation between the velocity and density perturbations reads $%
\rho _{1}=(\rho _{0}/\sin \theta )\partial v_{y}/\partial \xi $, i.e.\ Eq.\ (%
\ref{eq:Alf-NL1}) reduces to 
\begin{eqnarray}
  &&
  \left[ \frac{\partial }{\partial t}+c_{A,\mathrm{sp}}\cos \theta \frac{%
  \partial }{\partial \xi }+\frac{c_{A}^{3}\cos ^{3}\theta }{2\omega
  _{c}^{2}\sin ^{2}\theta }\frac{\partial ^{3}}{\partial \xi ^{3}}\right]
  v_{y}
  \nonumber \\ && \qquad
  = - \frac{(\mu _{B}B_{0})^2}{2mT}\frac{v_{y}^{2}}{%
  c_{A}^{3}}\frac{\partial v_{y}}{\partial \xi }  \label{eq:ALf-NL2}
\end{eqnarray}
which is a modified Korteweg de Vries (MKdV) equation with a focusing type
of nonlinearity. As is wellknown (see e.g.\ Ref.\ \cite{Cooney1993}), this
equation admits sech-shaped soliton solutions where the product of the
amplitude and the width is a constant, according to 
\begin{eqnarray}
  &&
  v_{y}=v_{y0}\,\mathrm{sech}\Bigg\{ \left( \frac{(\mu _{B}B_{0}\omega
  _{c})^{2}v_{y0}^{2}\sin ^{2}\theta }{3mTc_{A}^{6}\cos ^{3}\theta }\right)
  ^{1/2}
  \nonumber \\ && \quad\times
  \left[ \xi -c_{A,\mathrm{sp}}\left( \cos \theta +\frac{(\mu
  _{B}B_{0})^2v_{y0}^{2}}{12mTc_{A}^{3}c_{A,\mathrm{sp}}}\right) t%
  \right] \Bigg\} .
\end{eqnarray}

Astrophysical environments may exhibit extreme fields. Neutron stars have
surface magnetic field strengths of the order of $10^{6}-10^{9}$ T \cite
{Beskin-book}, while magnetars field strengths can reach $10^{10}-10^{11}$ T 
\cite{magnetar}, coming close to energy densities corresponding to the
Schwinger limit \cite{marklund-shukla}; here, in the vicinity of magnetars,
the quantum vacuum becomes fully nonlinear. However, more moderate, but
still very strong, fields appear at a distance from the surface of
magnetized stars, and often in conjunction with a pair plasma due to
cascading processes \cite{asseo}. 
Let us therefore consider a specific example of solitons in a pulsar
environment. We then take the unperturbed magnetic field as $B_{0}\simeq
10^{7}\,\mathrm{T}$. For a pulsar period of $P=1\,\mathrm{s}$, and a
multiplicity $n=10$ \cite{luo-etal}, the Julian-Goldreich expression $n_{%
\mathrm{JG}}=7\times 10^{15}(0.1\,\mathrm{s}/P)(B/10^{8}\,\mathrm{T})\,%
\mathrm{m}^{-3}$ for the pair plasma density gives $\rho _{0}\simeq
10^{-15}\,\mathrm{kg/m}^{3}$ \cite{asseo}. Furthermore, as  supported by the
work of e.g. Ref. \cite{Moskalenko1998}, we let the pair plasma temperature
be moderately relativistic, i.e. $T\simeq 0.4mc^{2}$. Finally, we let the
Alfv\'{e}n waves propagate almost parallell to the magnetic field, at an
angle $\theta \simeq 0.2\,\mathrm{rad}$, and have a velocity amplitude $%
v_{y0}\simeq 10^{3}\,\mathrm{m/s}$. We then find that the width of the
soliton is $d=[(\mu _{B}B_{0}\omega _{c})^{2}v_{y0}^{2}\sin ^{2}\theta
/3mTc_{A}^{6}\cos ^{3}\theta ]^{-1/2}\simeq  10^{-2}\,\mathrm{m}$. We
note that variations of the parameters in the expression for the soliton
width may change the size significantly, but this may also require other
modifications to be made, such as the inclusion of relativistic effects 
\footnote{Using the non-relativistic Pauli equation means
that certain effects are omitted, such as 
the spin-orbit coupling. Furthermore, a moderately relativistic
temperature means that a relativistic pressure model would be preferable.
However, since the waves considered are only weakly compressional (as
associated with the weak despersion of the Alfv\'{e}n waves), the main role
of the thermal effects is to determine the average orientation of the spins.
Thus it is not crucial to use a relativistic pressure model in the example given here. 
Finally, for the low amplitude solitons considered, there is
clearly no need to account for a relativistic quiver velocity.}. 
However, at this stage, we conclude that the existence
of localized spin structures in the pair plasmas surrounding pulsars is
likely, and further generalizations are left for future studies.


\section{Summary and Discussion}


In the present paper we have studied the dynamics of an electron- positron
pair plasma in the MHD limit, and included the spin properties of the
constituents. A closed set of one-fluid equations have been obtained,
resembling the standard MHD equations but including both a standard quantum
force (from the so called Bohm potential), as well as a number of new terms
related to the particle spins. The spin terms are of particular importance
for strongly magnetized plasmas and for low temperature plasmas, when the
spins are aligned with the magnetic field. We stress that the spin terms can
have rather different properties than the usual terms in the MHD equations.
As a consequence, it turns out that the spin force can be of importance even
when its magnitude is smaller than the usual $\mathbf{j}\times \mathbf{B}$- force.
In order to demonstrate this property, we have studied the special case of
weakly nonlinear shear Alfv\'{e}n waves in the one-dimensional limit. Due to
the spin effects, such waves may be governed by a MKDV-equation, leading to
wave steepening and subsequent soliton formation. By contrast, the
nonlinearity cancels to all orders for such waves, if the spin terms are
omitted. The results of the present paper can be of particular singnificance
for astrophysical pair plasmas in the vicinity of pulsars and magnetars \cite
{baring-etal,harding-lai}, as well as low-temperature laboratory plasmas \cite{rydberg}, and 
nano-structured systems \cite{ferro}. 

\acknowledgments

The authors would like to thank P. K. Shukla and L. Stenflo for stimulating
discussions. This research was supported by the Swedish Research Council.

\end{document}